\documentclass{jpp}
\usepackage{graphicx,natbib}

\ifCUPmtlplainloaded \else
  \checkfont{eurm10}
  \iffontfound
    \IfFileExists{upmath.sty}
      {\typeout{^^JFound AMS Euler Roman fonts on the system,
                   using the 'upmath' package.^^J}%
       \usepackage{upmath}}
      {\typeout{^^JFound AMS Euler Roman fonts on the system, but you
                   dont seem to have the}%
       \typeout{'upmath' package installed. JPP.cls can take advantage
                 of these fonts, if you use 'upmath' package.^^J}%
      }
  \else
  \fi
\fi

\usepackage{amsmath,amssymb}

\begin{document}

\title{Electron mirror instability: Particle-in-cell simulations}

\author{Petr Hellinger and \v St\v ep\'an \v Stver\'ak}

\affiliation{Astronomical Institute, CAS,
Bocni II/1401, CZ-14000 Prague, Czech Republic\\[\affilskip]
Institute of Atmospheric Physics, CAS,
Bocni II/1401, CZ-14000 Prague, Czech Republic
}

\pubyear{2018}
\volume{?}
\pagerange{??}
\date{??}

\maketitle

\begin{abstract}
Properties of the electron mirror instability
and its competition with the usually dominant whistler (electron cyclotron)
instability driven by the electron perpendicular
temperature anisotropy
are investigated on the linear level using a Vlasov linear solver and
on the nonlinear level using a two-dimensional full particle code.
The simulation results show that the linearly subdominant
electron mirror instability may compete on the nonlinear level 
with the whistler instability and may even become eventually the dominant mode
that generates robust non-propagating sub-ion-scale coherent structures in the form of magnetic peaks.
\end{abstract}

\section{Introduction}

Large scale motion of magnetised plasmas may lead to variation of
the magnitude of the magnetic field and of the plasma density.
In the collisionless limit these changes drive particle temperature
anisotropies that may lead to various plasma instabilities \citep{gary93}.
 Here we consider a homogeneous collisionless magnetised plasma 
consisting of electrons and protons.
In this case, the perpendicular electron temperature anisotropy (i.e.
when the perpendicular temperature is larger than the parallel one) in 
plasmas may generate (at least) two electromagnetic instabilities:
the whistler/electron cyclotron instability that
drives the whistler waves with the most unstable 
modes at electron scales parallel propagating with respect to the ambient
magnetic field 
\cite[except at very low beta plasmas, where the most unstable mode
shifts to oblique angles, cf.][]{garyal11}
and the mirror mode with wave vectors oblique with respect to the ambient
magnetic field.
The mirror instability driven only by anisotropic electrons is sometimes called 
the field-swelling instability \cite[cf.][]{baco84,migl86}. 
Similarly, the perpendicular  proton temperature anisotropy
may generate (at least) two electromagnetic instabilities:  
the (parallel) proton cyclotron and the (oblique) mirror instability.
Near threshold the mirror instability has only one branch,
 the most unstable mode appears at scales much larger
than the characteristic particle scales (inertial lengths and gyroradii)
and at strongly oblique angles with respect to the ambient magnetic field
and both the species contribute to its destabilization.
Further away from threshold the most unstable
modes get shorter length scales and less oblique angles
and the mirror instability may have two separate maxima
one at larger (proton) scales and one at shorter (electron) scales \citep{noreal17}. 
When the most unstable mode 
appears on proton/electron scales the instability
will be called the proton/electron mirror instability here.
We note, however, that the distinction between the two (and
the large-scale mirror instability) is
somewhat arbitrary, similarly to the distinction between
proton and electron (and large) scales. On the other hand, the spatial
and temporal scales
of fluctuations determine the properties of their interaction with particles.

Mirror mode waves/structures are observed in the solar wind,
planetary magnetosheaths and plasma sheets 
\citep{wintal95,joyal06,tsural11,enrial13}. These structures are non-propagating and
pressure-balanced and have a form of magnetic enhacements/peaks or
depressions/holes with scales larger than
but comparable to
ion gyroscales. Magnetic peaks are typically observed in the region
unstable with respect to the mirror instability whereas 
magnetic holes are seen in the stable region
\citep{soucal08,genoal09}.
Isolated magnetic holes with a wide range of scales are often observed in the solar wind
and some of them are likely related to the mirror instability \citep{stka07}.
Sub-ion-scale magnetic peaks \citep{yaoal18} and 
 holes \citep{geal11,yaoal17,zhanal17} are also sometimes observed.
These structures are possibly generated by the electron mirror instability.
However,
the competing whistler instability driven 
by the electron perpendicular temperature anisotropy generally dominates over
the electron mirror instability
\citep{gaka06} even for relatively large electron betas.  
It is therefore questionable if the electron mirror instability is
a relevant process in the space plasma context, in particular, 
 for generating sub-ion-scale magnetic holes; other
mechanisms may generate such structures \citep{balial12,sundal15}.
On the other hand,
the linear predictions are usually obtained assuming bi-Maxwellian particle
distribution functions whereas in reality the electron distributions observed
in situ 
are very different from bi-Maxwellian ones and generally include
an important population of non-thermal particles with a  power-law-like 
distribution \citep{stveal09} that affect the linear plasma properties \citep{pieral16,shaaal18}.
Furthermore, a competition between different instabilities
has to be investigated at the nonlinear level. Here we use
a two-dimensional (2-D) version of a full particle-in-cell (PIC) code
starting from an initial condition where both the whistler and
electron mirror instability are unstable to see their competition.
This paper is organized as follows: section~\ref{linear} presents linear
predictions for the two instabilities,
section~\ref{simulations} 
describes the numerical code and
presents the simulation results.
The presented results are summarized and discussed in section~\ref{discussion}.

\section{Linear predictions}
\label{linear}
The condition for the mirror instability
in a plasma consisting of an arbitrary set of species (denoted by the subscript s) with
  bi-Maxwellian velocity distribution functions non-drifting with respect to each other
can be given as $\Gamma>0$ where
\begin{equation}
\Gamma=\sum\limits_{\mathrm{s}} \beta_{\perp\mathrm{s}} 
\left(\frac{T_{\perp \mathrm{s}}}{T_{\|\mathrm{s} }} -1\right)
- 1 - \frac{
\left(\sum_{\mathrm{s}}\rho_{\mathrm{s}} {T_{\perp \mathrm{s}}}/{T_{\|\mathrm{s} }}\right)^2 }
      {2\sum_{\mathrm{s}} {\rho_{\mathrm{s}}^2}/{\beta_{\| \mathrm{s}}}} 
\label{crit}
\end{equation}
where $T_{\perp \mathrm{s}}$ and $T_{\|\mathrm{s} }$ denote the perpendicular
and parallel particle temperatures, respectively, $\beta_{\perp\mathrm{s}}$ and $\beta_{\|\mathrm{s}}$ 
denote the perpendicular
and parallel betas (ratios between the particle and magnetic pressures), respectively, 
$\rho_{\mathrm{s}}$ denotes the particle charge densities; $\sum_{\mathrm{s}}$ denotes a
sum over all species
\cite[cf.][]{stix62,hase69,hell07}; here we consider only protons and electrons with the subscripts p and e, respectively.

\begin{figure*}
\includegraphics[width=32pc]{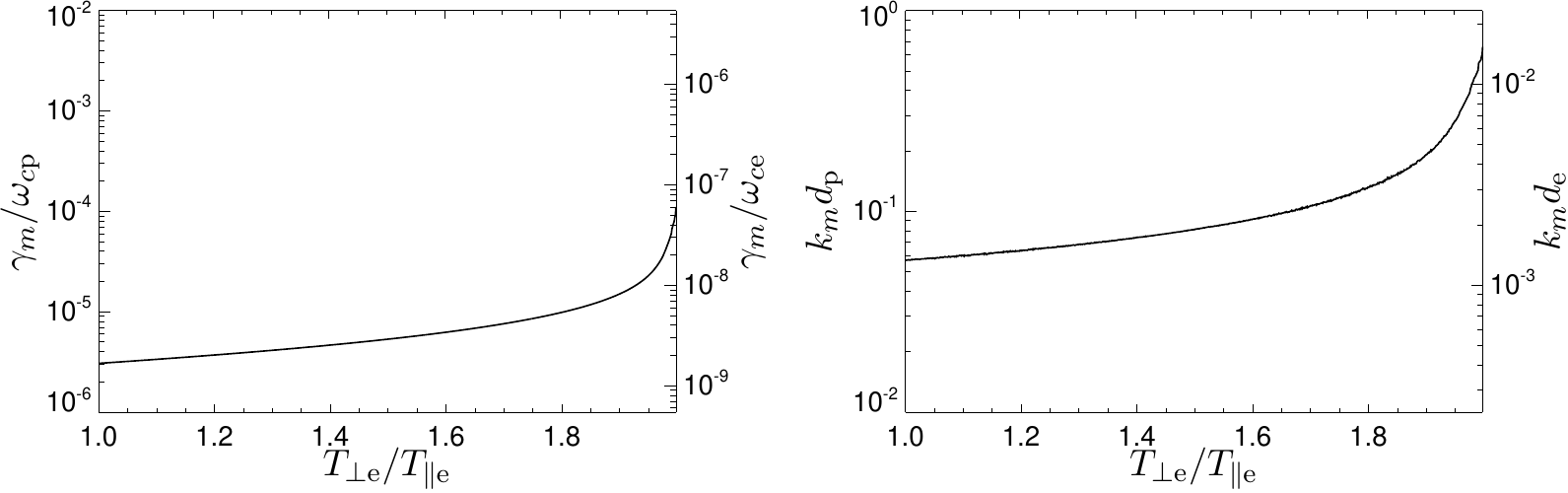}
\caption{(left) The maximum growth rate $\gamma_m$ of the
 mirror instability (normalized to the proton cyclotron frequency  $\omega_{c\mathrm{p}}$ or
to the electron one $\omega_{c\mathrm{e}}$)
and (right) the wave vector of the most unstable mode $k_m$
(normalized to the proton inertial length $d_{\mathrm{p}}$ or to the electron one $d_{\mathrm{e}}$)
along a one-dimensional subspace of the four dimensional space
$(\beta_{\|\mathrm{p}},T_{\perp \mathrm{p}}/T_{\|\mathrm{p}},\beta_{\|\mathrm{e}},T_{\perp \mathrm{e}}/T_{\|\mathrm{e}})$
determined by the three conditions $\Gamma=0.01$, $\beta_\mathrm{e}=\beta_\mathrm{p}=1$ (see the text).
This subspace is parametrized by
$T_{\perp \mathrm{e}}/T_{\|\mathrm{e}}$.
\label{epmir1} }
\end{figure*}

\begin{figure*}
\includegraphics[width=32pc]{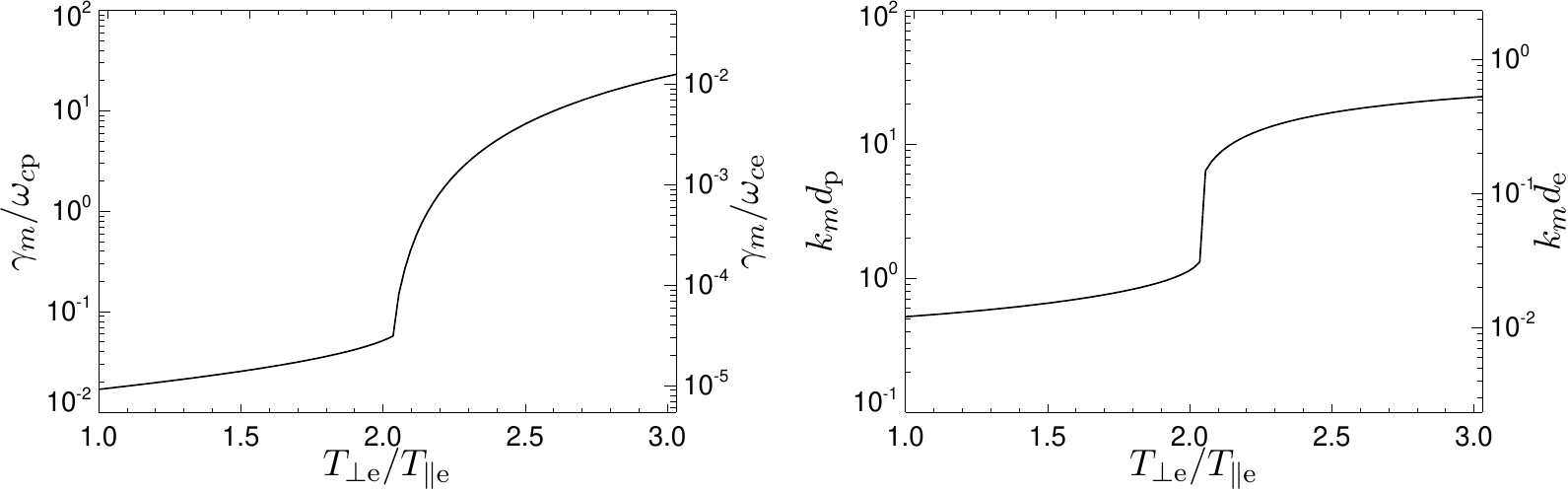}
\caption{(left) The maximum growth rate $\gamma_m$ of the
electron/proton mirror instability 
and (right) the wave vector of the most unstable mode $k_m$
along a one-dimensional subspace of the four dimensional space
$(\beta_{\|\mathrm{p}},T_{\perp \mathrm{p}}/T_{\|\mathrm{p}},\beta_{\|\mathrm{e}},T_{\perp \mathrm{e}}/T_{\|\mathrm{e}})$
determined by the three conditions $\Gamma=1$, $\beta_\mathrm{e}=\beta_\mathrm{p}=1$ (see the text).
This subspace is parametrized by
$T_{\perp \mathrm{e}}/T_{\|\mathrm{e}}$.
\label{epmir} }
\end{figure*}

We start with a short analysis of the relationship
between the proton and electron mirror instability far from threshold. 
First, we investigate
the linear prediction for the mirror instability near threshold for $\Gamma=0.01$ (i.e., we keep the same distance
from threshold)
and for constant total electron and proton betas,
$\beta_\mathrm{e} =\beta_{\|\mathrm{e}}(1+2 T_{\perp \mathrm{e}}/T_{\|\mathrm{e}})/3=1$ and
$\beta_\mathrm{p}= \beta_{\|\mathrm{p}}(1+2 T_{\perp \mathrm{p}}/T_{\|\mathrm{p}})/3=1$,
using a full linear Vlasov solver \citep{hellal06}.
The system constrained by $\Gamma=1$ and $\beta_\mathrm{e}=\beta_\mathrm{p}=1$
 gives a one-dimensional dependence of the maximum growth rate $\gamma_m$
on the electron temperature anisotropy ($T_{\perp \mathrm{p}}/T_{\|\mathrm{p}}$
decreases while $T_{\perp \mathrm{e}}/T_{\|\mathrm{e}}$ increases to keep $\Gamma$ constant;
the maximum of $T_{\perp \mathrm{e}}/T_{\|\mathrm{e}}$ corresponds to isotropic protons
and vice versa). 
The results of this calculation
are shown in Figure~\ref{epmir1}. This figure shows that 
the maximum growth rate $\gamma_m$ is small fraction of the proton cyclotron
frequency, $\omega_{c\mathrm{p}}$,
and increases with the increasing electron temperature anisotropy. 
The wave length of the most unstable mode $2\pi/k_m$ is relatively large with respect 
to the proton inertial length $d_\mathrm{p}$ (here comparable to the proton gyroradius $\rho_\mathrm{p}$) and increases with $T_{\perp \mathrm{e}}/T_{\|\mathrm{e}}$ and remains large with respect to the electron inertial length $d_\mathrm{e}$. These results indicate
that even close to threshold the typical spatial temporal scales of the
mirror mode are influenced by the particles with a dominant perpendicular temperature
anisotropy.

The same analysis for $\Gamma=1$, i.e. relatively far from threshold 
is shown in Figure~\ref{epmir}.
In the case, when the proton temperature anisotropy dominates, $\gamma_m$ is
a small fraction of $\omega_{c\mathrm{p}}$
(with $k_m$ on proton scales) whereas
when the electron temperature anisotropy becomes dominant, there
is a rapid transition and $\gamma_m$ is then rather
a small fraction of the electron cyclotron frequency $\omega_{c\mathrm{e}}$
(with $k_m$ on electron scales) and comparable or larger than $\omega_{c\mathrm{p}}$. 
In the transition region there may exist two separate local maxima
\citep{noreal17}. The behaviour seen in Figure~\ref{epmir} is quite
general, we observe similar properties for $\beta_\mathrm{e}=\beta_\mathrm{p}=10$,
$\beta_\mathrm{e}=10$ and $\beta_\mathrm{p}=1$,
and $\beta_\mathrm{e}=1$ and $\beta_\mathrm{p}=10$ while keeping $\Gamma=1$.
Henceforth, we assume that protons are isotropic and the perpendicular
electron temperature anisotropy is only the only source of free 
energy for instabilities.

\begin{figure*}
\includegraphics[width=32pc]{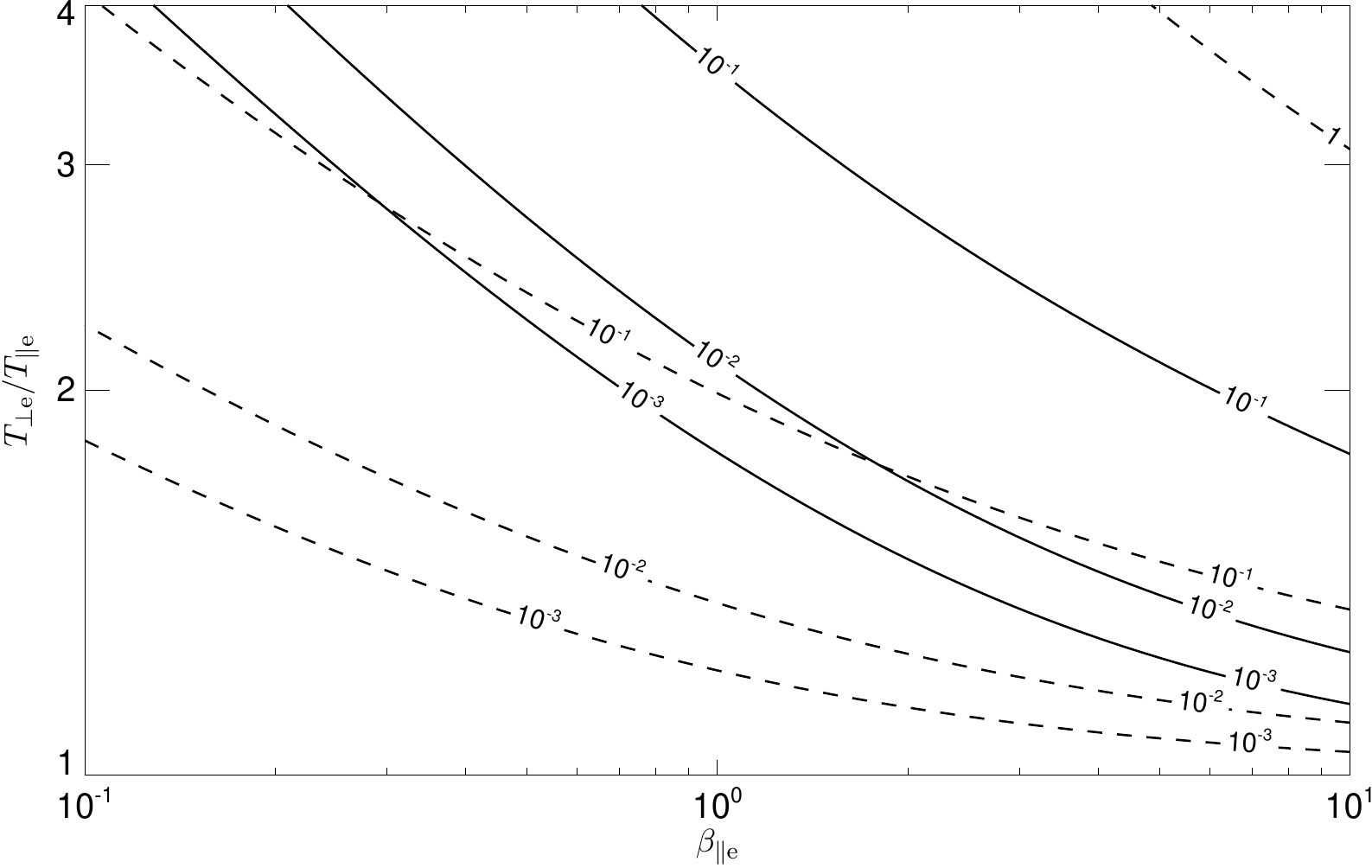}
\caption{The maximum growth rate $\gamma_m$ of (solid)
the mirror instability and (dashed) the electron whistler instability as a
function of the electron beta and the electron
temperature anisotropy. The solid and dashed contours are annotated
by $\gamma_m$ normalized to $\omega_{c\mathrm{e}}$.}
\label{emir}
\end{figure*}

Figure~\ref{emir} shows the maximum growth 
rate $\gamma_m$
 of (solid) the electron mirror and (dashed) the whistler instability  as a  function
of the electron beta and the electron
temperature anisotropy assuming bi-Maxwellian electrons
and Maxwellian protons; the ratio between the electron 
plasma and cyclotron frequencies is assumed to be 
$\omega_{p\mathrm{e}}/\omega_{c\mathrm{e}}=100$ (relevant for the solar wind).

The comparison between the linear predictions for the whistler and mirror instabilities in
Figure~\ref{emir} clearly confirms that the whistler instability is typically linearly dominant
and one expects that the electron mirror instability is not generally very relevant
\citep{gaka06}.
However, the linear predictions have limited applicability, it is necessary to
take into account the nonlinear
behaviour of the two instabilities.

\section{Simulation results}
\label{simulations}

To investigate nonlinear properties of the two instabilities and their competition we use 
a 2-D version of an explicit electromagnetic PIC code that employs the Darwin
approximation \citep{decy07,schral10}.
 The Darwin PIC model neglects the transverse
 component of the displacement current (but keeps the longitudinal part)
in the full set of Maxwell's equations, which makes them radiation-free,
but leaves the whistler physics unaffected from its fully electromagnetic
 counterpart \citep{hewe85};
The radiation-free Darwin approximation removes
the demanding Courant-Friedrich-Levy condition for a time step given by the speed of light \citep{schral10,hellal14}.
Consequently, the time step is set
by the greater of the electron plasma frequency or the electron cyclotron frequency.

Here we use the real mass ratio $m_\mathrm{p}/m_\mathrm{e}=1836$.
For the frequency ratio we use $\omega_{p\mathrm{e}}/\omega_{c\mathrm{e}}=4$
in order to save the numerical resources (by reducing the ratio
between the electron inertial length $d_{\mathrm{e}}$ and the Debye length $\lambda_D$).
Note that the linear prediction in Figure~\ref{emir} is calculated for $\omega_{p\mathrm{e}}/\omega_{c\mathrm{e}}=100$;
further linear analysis shows that the 
linear properties of the electron whistler and mirror instabilities depend only weakly 
on this ratio, a calculation analogous to  Figure~\ref{emir} for  $\omega_{p\mathrm{e}}/\omega_{c\mathrm{e}}=4$
gives results almost identical to that of Figure~\ref{emir}.

The electrons have initially $\beta_{\|\mathrm{e}}=10$
and $T_{\perp\mathrm{e}}/T_{\|\mathrm{e}}= 1.52$ whereas
the protons are initially isotropic with $T_\mathrm{p}=T_{\|\mathrm{e}}$.
The simulation box is chosen to be
a 2-D grid $ 2048^2$ with the physical sizes
 $\sim 1145^2 d_{\mathrm{e}}^2$.
The magnetic field is chosen to be along the $x$-direction.
There are $1024$ macroparticles per cell for electrons
and $512$  macroparticles per cell for protons.
The time step is $\Delta t=0.05 \omega_{c\mathrm{e}}^{-1}$.

\begin{figure*}
\includegraphics[width=32pc]{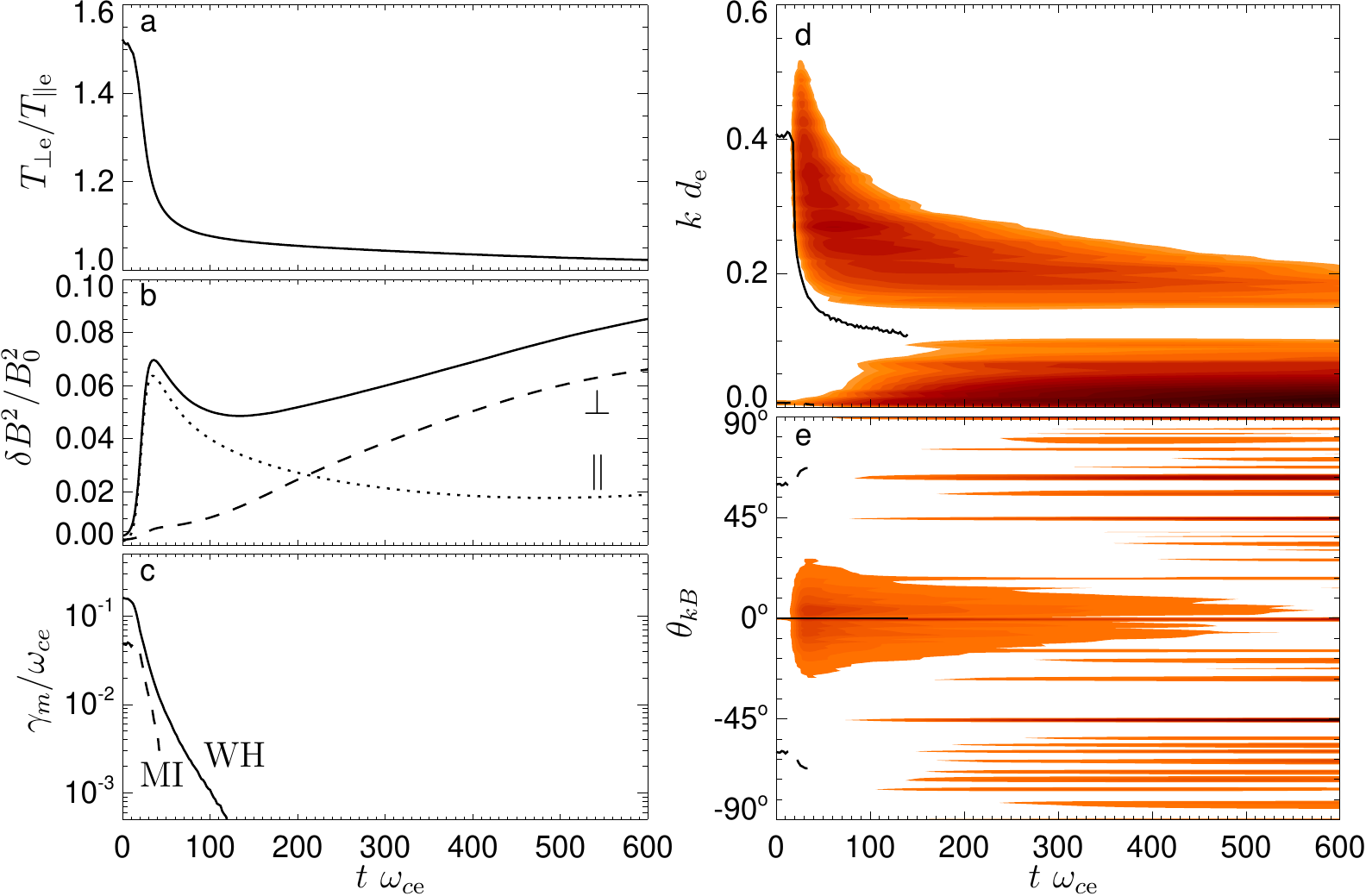}
\caption{
(left)
The electron temperature anisotropy $T_{\perp\mathrm{e}}/T_{\|\mathrm{e}}$ (a),
the fluctuating magnetic field $\delta B^2/B_0^2$ (b), and
the instantaneous maximum growth rate $\gamma_m$ normalized to $\omega_{c\mathrm{e}}$
 (c) as functions of time. 
Dotted and dashed line on panel (b) shows the fluctuating magnetic field
with $\theta_{kB}\le 30^\mathrm{o}$ and $\theta_{kB}> 30^\mathrm{o}$, respectively.
(right) Color scale plots of
the fluctuating magnetic field $\delta B$ as a function of time and wave vector $k$
(d) and as a function of time and propagation angle $\theta_{kB}$
(e). Solid and dashed lines (on panels d and e) show the properties ($k$ and $\theta_{kB}$)
of the most unstable mode for the whistler and mirror instability (see panel c), respectively.
}
\label{db}
\end{figure*}

The time evolution of the system is shown in Figure~\ref{db}:
Figure~\ref{db}a shows the evolution of the electron temperature anisotropy $T_{\perp\mathrm{e}}/T_{\|\mathrm{e}}$.
Figure~\ref{db}b displays the total fluctuating magnetic field $\delta B^2/B_0^2$ (solid line) as a function of time;
for comparison,
the dotted and dashed curves display the fluctuating magnetic field with quasi-parallel ($\theta_{kB}\le 45^\mathrm{o}$)
and quasi-perpendicular ($\theta_{kB}>45^\mathrm{o}$) wave vectors, respectively.
Figure~\ref{db}c presents results of the linear analysis based on the instantaneous 
electron velocity distribution function \citep{hetr11,hellal14}, the maximum
growth rate $\gamma_m$ for the whistler (solid) and mirror (dashed) instabilities as
functions of time.
Figures~\ref{db}d and \ref{db}e  display color  scale plots of
the fluctuating magnetic field $\delta B$ as a function of time and wave vector $k$
 and as a function of time and propagation angle $\theta_{kB}$, respectively.
Solid and dashed lines (on panels d and e) show the properties ($k$ and $\theta_{kB}$)
of the most unstable mode for the whistler and mirror instability (see panel c), respectively.

Figure~\ref{db} shows that the plasma system with anisotropic electrons generates mostly quasi-parallel
(whistler) waves at electron scales that importantly reduce the electron temperature anisotropy.
At the same time, at relatively large scales (and oblique propagation) other fluctuations appear.
The amplitude of the whistler waves decreases after the saturation (at about $40/ \omega_{c\mathrm{e}}$).
The (quasi-)linear prediction based on the instantaneous electron distribution function 
indicate that the whistler mode is stabilized after $t \simeq 100/\omega_{c\mathrm{e}}$;
the electron mirror instability is linearly stabilized earlier, after $t \simeq 50/ \omega_{c\mathrm{e}}$.
The evolution of whistler wave activity follows to some extent the quasi-linear prediction (solid lines
on Figures~\ref{db}d and \ref{db}e). The
spectrum of magnetic fluctuations shifts to large scales after the saturation in agreement
with the shift of the most unstable mode (and also the whole unstable region) to larger scales
a short time before the saturation (the natural delay is owing to the fact the growth/damping
rate gives just the relative rate of amplitude change). The initial decrease of the whistler wave 
activity seems to be related to the shift of the unstable region, some short-wavelength modes that reached
important amplitudes become damped.

\begin{figure*}
\includegraphics[width=32pc]{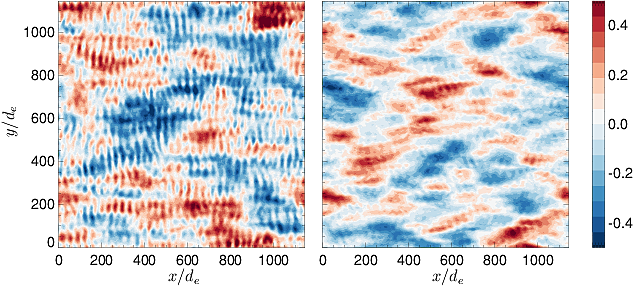}
\caption{
Color scale plots of 
the fluctuating magnetic field (left) $\delta B_z$ and (right) $\delta B_x$  as a function of $x$ and $y$
at the end of the simulation.
}
\label{bxz}
\end{figure*}

In contrast with the (quasi-)linear prediction, the simulation exhibits a continuous increase of
fluctuations at oblique angles. These fluctuations have a form of coherent structures as
seen in
Figure~\ref{bxz}. This figure displays the spatial structure of the fluctuating magnetic field 
($\delta B_z$ and $\delta B_x$ components)
at end of the simulation. At this time the system is quasi-stationary, 
we don't observe big/qualitative changes during the period 
$500$--$600\omega_{c\mathrm{e}}^{-1}$.
Figure~\ref{bxz} shows that the electromagnetic fluctuations are composed of
quasi-parallel (whistler) waves and oblique compressible structures.
These structures are essentially non-propagating and exhibit a weak anti-correlation
between $B_x$ and the electron density $n_e$.

\begin{figure*}
\includegraphics[width=26pc]{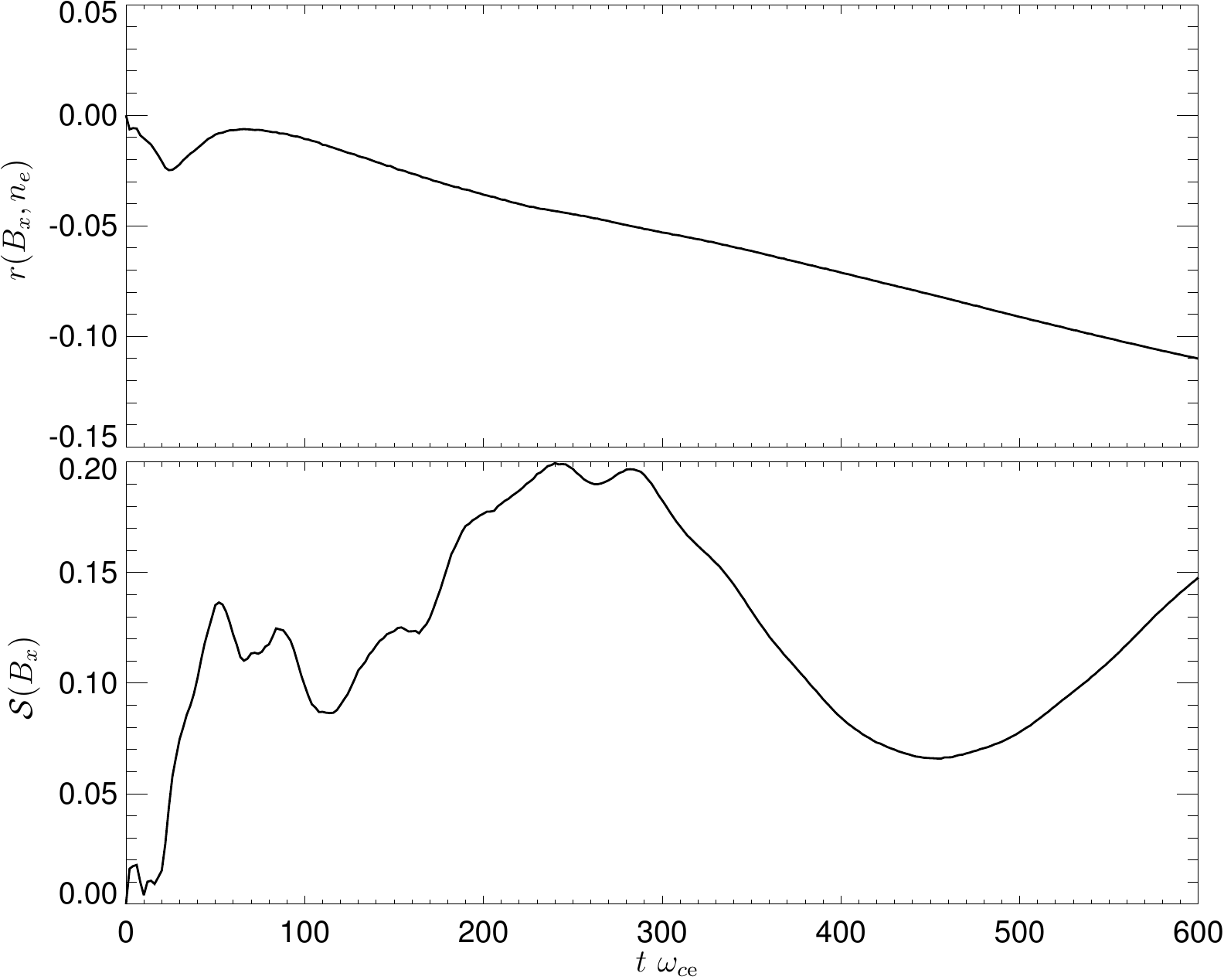}
\caption{
Evolution of the system, (top) the correlation coefficient (solid) between the compressible magnetic
component $B_x$ and the electron number density $r( B_x, n_e)$,
(bottom)
and the skewness of $B_x$  $\mathcal{S}(B_x)$ as functions of time.
}
\label{corskew}
\end{figure*}

Figure~\ref{corskew} shows the evolution of the system, the correlation coefficient between the compressible magnetic
component $B_x$ and the electron number density $r ( B_x, n_e)$
and the skewness of $B_x$  $\mathcal{S}(B_x)$ as functions of time.
The (sample) correlation coefficient of two discrete variables $x$ and $y$ is given by
\begin{equation}
r(x,y)= \frac{ \frac{1}{n} \sum_{i=1}^{n} (x_i-\bar{x})(y_i-\bar{y})  }{\sqrt{\sum_{i=1}^{n} (x_i-\bar{x})^2} \sqrt{\sum_{i=1}^{n} (y_i-\bar{y})^2}}
\end{equation}
where $\bar{x}= \sum_{i=1}^{n} x_i/n$ and $\bar{y}= \sum_{i=1}^{n} y_i/n$ are the mean values, $n$ is the number of grid points.
The (sample) skewness of a discrete variable $x$ is given by
\begin{equation}
\mathcal{S}(x)= \frac{\sqrt{n(n-1)}}{n-2} \frac{ \frac{1}{n} \sum_{i=1}^{n} (x_i-\bar{x})^3 }{\left[\frac{1}{n} \sum_{i=1}^{n} (x_i-\bar{x})^2 \right]^{3/2}}.
\end{equation}

The compressible magnetic component $B_x$ is weakly anti-correlated with the electron density 
through-out the simulation; this anti-correlation becomes more pronounced at later
times when the oblique fluctuations get stronger. This property is consistent
with the linear as well as nonlinear expectation for the electron mirror instability.
The distribution of values $B_x$ has a positive skewness through-out the simulation.
It grows, oscillates, decreases for $300/\omega_{c\mathrm{e}}\lesssim t \lesssim 450/\omega_{c\mathrm{e}}$, 
and grows again after $t\gtrsim450/\omega_{c\mathrm{e}}$. This indicates that the coherent structures
seen in the simulation (see Figure~\ref{bxz}) are rather magnetic enhancements (peaks/humps) \cite[cf.][]{genoal09}.
We interpret the oblique fluctuations as nonlinear electron mirror structures.

\begin{figure}
\includegraphics[width=32pc]{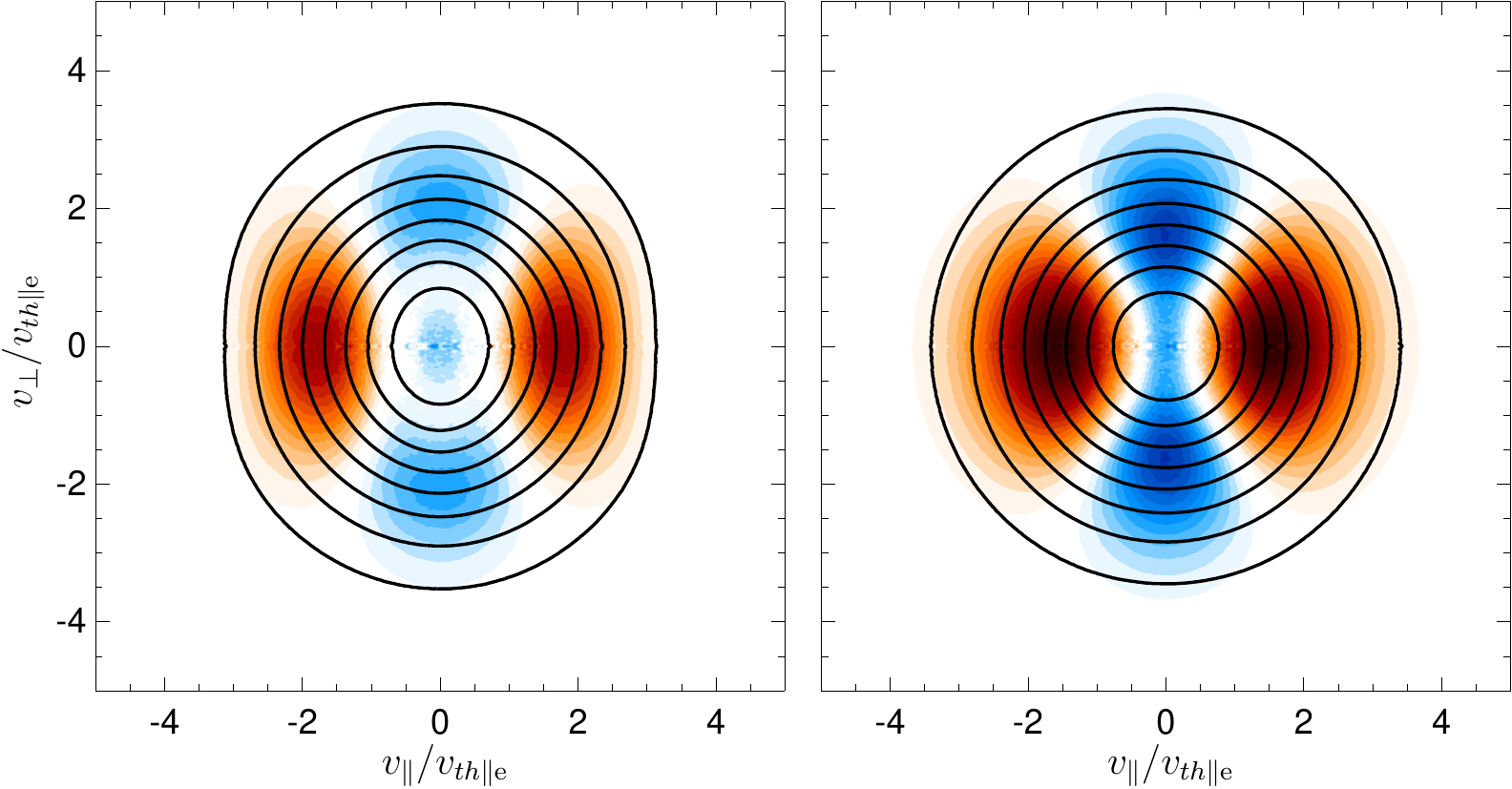}
\caption{
Electron velocity
distribution functions shown as solid contours at (left) $t=36\omega_{c\mathrm{e}}^{-1}$ and (right) $t=600\omega_{c\mathrm{e}}^{-1}$.
The color scale plots display the change of the distribution function $\delta f=f(t)-f(0)$
where shades of red (blue) denote positive (negative) values.
}
\label{vdf}
\end{figure}

Finally, Figure~\ref{vdf} shows the electron velocity
distribution functions at the time of the saturation of the whistler instability ($t=36\omega_{c\mathrm{e}}^{-1}$) and at the
end of the simulation ($t=600\omega_{c\mathrm{e}}^{-1}$). This figure suggests that around the whistler saturation the electron
distribution function have a shape reminiscent of a quasi-linear cyclotron plateau \citep{keen66}.
At later stages of the simulation, when the whistler activity decays and
electron mirror structures become dominant, the electron velocity distribution  
function has a shape that is relatively close to a bi-Maxwellian one with
a weak anisotropy (see Figure~\ref{db}). For the sake of completeness, we note that  
protons are essentially unaffected by the electron whistler and mirror modes,
the relative change of their temperature is of the order of $10^{-4}$
during the simulation.

\section{Discussion}
\label{discussion}

In a proton-electron plasma the mirror instability may have 
the most unstable mode on proton or electron scales
 far from threshold.
We investigated the linear prediction for one parameter case where
we kept the same distance from threshold and the same total betas for
protons and electrons. The most unstable mode appears on proton scales roughly
in the region where protons were more anisotropic then electrons
and vice versa.
How this linear prediction works in the four-dimensional
space of proton and electron temperature anisotropies and (parallel) betas
is an open problem beyond the scope of this paper; this behaviour is also
affected by the presence of suprathermal populations \citep{shaaal18}. These results
could also depend on the ratio $\omega_{p\mathrm{e}}/\omega_{c\mathrm{e}}$
but our results indicate that this dependence is weak.

2-D PIC simulation of a system initially unstable with
respect to both the whistler and electron
mirror instabilities shows that
the linearly dominant whistler instability rapidly generates
quasi-parallel whistler waves that reduce the electron temperature anisotropy.
Meanwhile, the mirror modes slowly grow but are rapidly linearly stabilized 
owing to the dominant whistler instability. However, the mirror modes
continue to grow and become dominant electromagnetic fluctuations
in the form of coherent magnetic enhancements/peaks on electron scales. At the same
time, the amplitude of whistler waves decreases probably partly owing
to the presence of important electron mirror structures; on the other hand, similar
decay is seen also without mirror modes for a relatively strong whistler
instability \citep{kimal17}.

A nonlinear growth of mirror structures that appears after the
(quasi-)linear saturation was observed in a numerical simulation (in the hybrid,
kinetic ion and fluid electron, approximation)
for the proton mirror instability \citep{calial08}; in this case
the proton mirror instability forms coherent structures in the form
of magnetic peaks on proton scales. We expect that the formation of electron mirror
structures seen in the full particle simulations is driven by a
similar/analogous phenomenon.
The nonlinear growth of the (proton) mirror structures
is in agreement with the nonlinear dynamic model of \cite{kuznal07a,kuznal07b}
for the mirror instability 
near threshold,  based on a reductive
perturbative expansion of the Vlasov-Maxwell equations.
This model extends the mirror dispersion relation by
including the dominant nonlinear coupling whose effect is to
reinforce the mirror instability. 
In the present case, the system is relatively far from threshold
but the simulations results are qualitatively
in agreement with this model. The
basic disagreement between the model and simulations,
seen also in the hybrid simulations, is that the model
predicts formation of magnetic depression/holes whereas
the simulations show formation of magnetic enhancements/peaks.
This difference may be related to the flattening of the distribution
function that may lead to a change of the nonlinear term \citep{hellal09}.

The present results are based on one case
with ad hoc parameters where  the electron temperature
anisotropy is quite strong so that
both the modes are initially unstable.
Consequently, the whistler instability has quite a fast
initial maximum growth rate. This may require a fast
anisotropization mechanism. On the other hand,
in a more realistic case where the temperature
anisotropy is continuously driven, for example
by the plasma expansion/compression 
\citep{hellal03,hellal03b,sina15,hell17a}
or by the velocity shear \citep{kunzal14,riqual15},
the dominant instability may not be efficient enough
to arrest the anisotropy above threshold of
the subordinate instability and the latter  may be
eventually destabilized and affect the system behaviour.
Driven simulations of \cite{traval07b} and \cite{ahmaal17} exhibit
formation of magnetic enhancements/peaks in the region 
unstable with respect to the proton mirror instability
and a transformation
of these peaks to magnetic holes as the system
becomes more stable in agreement with in situ observations 
\citep{genoal11}. We expect a
similar behaviour for the electron mirror instability,
 some of the observed
sub-ion-scale magnetic holes \citep{geal11,yaoal17,zhanal17}
may be generated by an analogous process.

The presented simulation results have many limitations.
We used a 2-D code. In the
 realistic three-dimensional case the oblique mirror modes
have more degree of freedom and likely play a more important role
\cite[cf.][]{shojal09}. 
The limited number of particles per cell leads (and the large beta)
leads to a non negligible noise level that affect the initial level
of fluctuation and, consequently, the nonlinear competition
of different modes.
Another problem is that
at the end of the simulation the amplitude of magnetic fluctuations is still
growing. The growth is, however, rather weak and we don't expect that the evolution
would change significantly at later times.
Also, we investigated a spatially homogeneous case. The evolution of
the system may be importantly influenced by the presence of important turbulent fluctuations.
However, the numerical results of \citep{hellal17} show that the (proton) mirror instability
can coexist with a developed strong turbulence (even in a constrained 2-D system)
so that we expect our results are relevant even for a turbulent plasma system.

In concluding, we showed using 2-D full PIC simulations that the linearly subdominant
electron mirror instability can efficiently compete on the nonlinear level with the whistler instability 
and even become eventually the dominant mode in the form of non-propagating coherent structures.
The coherent structure are relatively robust, persist (and even grow nonlinearly) in the region
stable with respect to the mirror instability and could have important effect on
electron transport in hot astrophysical plasmas \citep{komaal16,robeal18}.

\acknowledgments
The authors acknowledge grant 15-17490S of the
Czech Science Foundation.
The (reduced) simulation data are available at the
Virtual Mission Laboratory Portal (http://vilma.asu.cas.cz).

\end{document}